\documentclass{article}
\usepackage[english]{babel}
\usepackage[utf8x]{inputenc}
\usepackage{bm}
\usepackage{amsmath, amssymb}
\usepackage{graphicx}

\title{RatingScaleReduction package: stepwise rating scale item reduction without predictability loss}
\author{by Waldemar W. Koczkodaj and Alicja Wolny--Dominiak}

\begin{document}
\maketitle

\begin{abstract}
	This study presents an innovative method for reducing the number of rating scale items without predictability loss. The ``area under the receiver operator curve method'' (AUC ROC) is used to implement in the \textbf{RatingScaleReduction} package posted on CRAN  \footnote{https://cran.r-project.org/web/packages/RatingScaleReduction/index.html}. Several cases have been used to illustrate how the stepwise method has reduced the number of rating scale items (variables). \\
\end{abstract}


\section{Introduction}
\label{intro}

Rating scales (also called assessment scale and ''the scale'' in this study) are used to elicit data about quantitative entities. Often, predictability of rating scales (also called ``assessment scales'') can be improved.
Rating scales often use values ``1 to 10'' and some rating scales may have over 100 items (questions) to rate. Sometimes, the scale is called a \textit{survey} or a \textit{questionnaire}. A questionnaire is a tool for data collection while a survey may not necessarily be conducted by questionnaires since some surveys may be conducted by interviews or by data gathered from web pages. In fact, the main and very important distinction between the scale and both questionnaires and surveys is that the scale is used for assessment (as in ``the scale of disaster'') while questionnaires and surveys may be used to  only collect data. In other words, some kind of ``summation procedure'' must be provided for questionnaires or surveys to become rating scales.

The recent popularity of rating scales is due to various ``Customer Reviews'' on the Internet  where five stars are often used instead of ordinal numbers. However, the most important examples of rating scales are questionnaires used in examinations. We may risk a statement that the accelerated progress of granting academic degrees can be linked to a better use of rating scales.

Rating scales are predominantly used to express our subjective assessments such as ``on the scale 1 to 5 express your preference'': ``strongly agree to strongly disagree'' (with 3 as 'neutral'' preference). The importance of subjectivity processing probably inspired introduction of the idea of \textit{bounded rationality}, proposed by Herbert A. Simon (the Nobel Prize winner), as an alternative basis for the mathematical modeling of decision making. It is often expressed as ``good enough is perfect'' and gained popularity in the software industry where frequent updates are common.
Objective data are more commonly used in so called ``strict sciences'' while processing subjectivity is still under development. It is worth noticing that the objectivity is illusive. Often, the difference  between subjectivity and objectivity is a matter of an arbitrary decision. For example, an item listed for sale for, let us say, 100,000 monetary units, will be very likely sold for 99,999 of such units if such an offer is made. If so, one may also not resist accepting 99,998 monetary units and so on. Setting a limit (so called, ``the bottom line'') is often a highly subjective decision. Scales help in many cases, but a large number of items (questions) is often a discouragement for its use.

It seems that one of the first successful rating scale reduction (RSR) took place in \cite{VC1979}. The 17-item Hamilton Rating Scale for Depression (HAM-D17) was used to derive even a more reduced version (Ham-D7) with seven items. According to \cite{MBB2002}, ``The clinical utility of the HAM-D17 is hampered, in part, by the length of time required to administer the interview and by the lack of inter-rater reliability.''

\section{Heuristic algorithm}
\label{HA}

A heuristic is, in essence, a simplified method for solving a problem more quickly when well-established methods fail to find a sound algorithm. Usually, this is achieved by the ``good enough is perfect'' approach, mentioned in the introduction, as characterization of the heuristic solution. Heuristics are expected to produce a reasonable solution when a time frame and accuracy are a problem. The ``good enough''solution for an urgent problem is commonly practiced in computer science. It is usually not the best solution to our problem but it may still be of  great value. For example, the traveling salesman problem (TSP), often formulated as ``find the shortest possible route to visit each city exactly once and return to the origin city'', cannot be solved for 50 or more cities by verifying all possible combinations since the total number of such combinations would easily exceeds the number of atoms in the entire Universe. Using heuristics, we can solve TSP for millions of cities with the accuracy of a small fraction of 1\%.
Most heuristics produce results by themselves but many are used in conjunction with optimization algorithms to improve their efficiency (e.g., differential evolution).

In our case, the number of possible combinations for a rating scale with 100 items (which is not uncommon) is a ``cosmic number'' hence the complete search must be ruled out. Computing the area under the receiver characteristic curve for all items is the basis for our heuristic. Common sense dictates that the contribution of the individual items to the overall value of the area under the receiver characteristic curve needs to be somehow utilized. We have decided on a stepwise heuristic. Certainly, the results need to be verified and used only if the item reduction is substantial.

\section{The package description}

Rating scales are used to elicit data about qualitative entities (e.g., research collaboration). This study presents an innovative method for reducing the number of rating scale items without  predictability loss. The ``area under the receiver operator curve method'' (AUC ROC) is used. The presented method has reduced the number of rating scale items (variables) to 28.57\% (from 21 to 6) making over 70\% of collected data unnecessary.

Results have been verified by two methods of analysis: Graded Response Model (GRM) and Confirmatory Factor Analysis (CFA). GRM revealed that the new method differentiates observations of high and middle scores. CFA proved that the reliability of the rating scale has not deteriorated by the scale item reduction. Both statistical analysis evidenced usefulness of the AUC ROC reduction method.

Rating scales (also called assessment scale) are used to elicit data about quantitative entities. Often, the predictability of rating scales (also called ``assessment scales'') could be improved.
Rating scales often use values: ``1 to 10'' and some rating scales may have over 100 items (questions) to rate.
Other popular terms for rating scales are: \textit{survey} and \textit{questionnaire}, although a questionnaire is a method of data collection while survey may not necessarily be conducted by questionnaires. Some surveys may be conducted by interviews or by analyzing web pages. Rating itself is very popular on the Internet for ``Customer Reviews'' where five stars (e.g., Amazon.com) are often used instead of ordinal numbers. One may regard such rating as a one item rating scale.

In computer science and mathematical optimization, a heuristic is a technique designed for solving
a problem for finding an approximate solution when classic methods fail to find any exact solution.
Often, finding such methods is achieved by trading completeness, accuracy, or optimality, for speed.

The main objective of a heuristic is to produce a solution that is good enough to solve our problem. The solution may not be the ``best'' solution and it may only approximate the solution since the optimal solution may require a prohibitively long time. The traveling salesman problem and virus scanning are probably the most recognized problems where the need for using heuristics is evident.
In both cases, the complete search for the optimal search would take thousands of years using the fastest computers built.
One of the shortest heuristics may be ``22/7'' as an approximation of $\Pi$ constant
with two decimal points (3.14), as it is easier to remember and sometimes easier to use.

Herbert A. Simon originally was the proponent of bounded rationality.
In practice, it means that human judgments are based on heuristics. He is the only person who received both the Nobel prize and the Turing prize.

Data collected by a rating scale with a fixed number of items (questions) are stored in a table with one decision (in our case, binary) variable. The parametrized classifier is usually created by total score of all items. The outcome of such rating scales is usually compared to external validation provided by assessing professionals (e.g., grant application committees).

Our approach not only reduces the number of items, but also sequences them according to the contribution to predictability. It is based on the Receiver Operator Characteristic (ROC), which gives individual scores for all examined items.
The term ``receiver operating characteristic'' (ROC), or ``ROC curve'' was coined for a graphical plot illustrating the performance of radar operators (hence ``operating''). A binary classifier represented absence or presence of an enemy aircraft. It was used to plot the fraction of true positives out of the total actual positives (TPR = true positive rate) vs. the fraction of false positives out of the total actual negatives (FPR = false positive rate).
Positive instances (P) and negative instances (N) for some condition are computed and stored as four outcomes of a 2 contingency table or confusion matrix, as follows:

\begin{table}[htbp]
	\centering
	\caption{The confusion matrix}
	\begin{tabular}{|c|c|}
		\hline
		\textbf{True Positives} & \textbf{False Positives} \\
		\hline
		\textbf{False Negative} & \textbf{True Negative} \\
		\hline
	\end{tabular}%
	\label{tab:c-m}%
\end{table}%

Each patient either has or does not have the disorder.
The screening outcome can be positive (classifying a patient as having the disorder)
or negative (classifying the patent as not having the disorder).
The screening results for each patient may or may not match the subject's actual status.

In means that in the medical terminology we may have:
\begin{itemize}
\item TP = true positive: patient is correctly identified as having the disorder,
\item FP = false positive: patient is incorrectly identified as having the disorder,
\item TN = true negative: patient with no disorder is correctly identified as not having the  disorder,
\item FN = false negative: patent with the disorder is incorrectly identified as not having the disorder.
\end{itemize}

\noindent In simple terms, positive = identified and negative = rejected hence: \\

\noindent True positive = correctly identified examples\\
False positive = incorrectly identified examples\\
True negative = correctly rejected examples\\
False negative = incorrectly rejected examples\\

In assessment and evaluation research, the ROC curve is
a representation of a ``separator'' (or decision) variable. The decision variable is usually: ``has a property'' or ``does not have a property'' or has some condition to meet (pass/fail).
The frequencies of positive and negative cases of the diagnostic test vary for the ``cut-off'' value for the positivity. By changing the ``cut-off'' value from 0 (all negatives) to a maximum value (all positives), we obtain
the ROC by plotting TPR (true positive rate also called sensitivity) versus FPR (false positive also called specificity) across varying cut-offs, which generate a curve in the unit square called an ROC curve.

According to \cite{TF2004}, the area under the curve (the AUC or AUROC) is equal to the probability that a classifier will rank a randomly chosen positive instance higher than a randomly chosen negative one (assuming the 'positive' rank higher than 'negative').

This can be seen as follows: the area under the curve is given by (the integral boundaries are reversed as large T has a lower value on the x-axis)

$$A = \int_{\infty}^{-\infty} y(T) x'(T) dT = $$ $$ \int_{\infty}^{-\infty} TPR(T) FPR'(T) dT = $$ $$\int_{-\infty}^{\infty} TPR(T) P_0(T) dT = \langle TPR \rangle$$

The angular brackets denote average from the distribution of negative samples.
AUC is closely related to the \textit{Mann-Whitney U test} which tests whether positives are ked higher than negatives. It is also equivalent to the Wilcoxon test of ranks.

ROC method is implemented by many R packages including: \textbf{pROC} \cite{pROC} and \textbf{ROCR} \cite{sing2005rocr}. There is also one interesting web application \textbf{easyROC} \cite{goksuluk2016easyroc} giving possibility to compute the confusion matrix and plot the curve on-line. The \textbf{RatingScaleReduction} package expands this analysis to carry out the procedure of rating scale reduction.\\

A package for preprocessing ``messy'' data into a form is easily analyzed within R is presented in  \cite{leeper2016crowdsourced}.
In \cite{winslow2016sbtools}, the new R package \textit{sbtools} enables users direct access to the advanced online data functionality provided by ScienceBase, the U.S. Geological Survey's online scientific data storage platform. It can be used for harvesting other data sets.

\section{Rating scale stepwise reduction procedure}
The procedure follows the heuristic algorithm represented by Fig.~\ref{figure:fig1}. Technically, it is an algorithm since the flowchart, represented by Fig.~\ref{figure:fig1}, shows the finite namer of steps. It is, however, a heuristic algorithms since the optimality of the presented approach cannot be guaranteed (as pointed out in Section~\ref{HA}). However, the common sense dictates to select the ``best'' attribute and keep adding to it the next ``best'' attribute where the ``best'' has the meaning of the area under the curve (AUC) value since it is the universally accepted criterion for classifiers (in statistical classification and machine learning). A rating scale total is a classifier. The classification in this study is regarded as an instance of supervised learning. Briefly, it requires a training set of correctly identified examples (observations) with the external evaluation. In our case, a trained professional is needed to determine if a subject (a screened psychiatric patient) had a mental disorder or not). An algorithm that implements a concrete classification is called as a classifier. The most common way of doing i by a rating scale is using the total of all items. Some of them may be negative (e.g., in the Oxford Happiness Questionnaire, see \cite{OHQ}).


\begin{figure}[htbp]
	\centering
	\includegraphics[scale=.5]{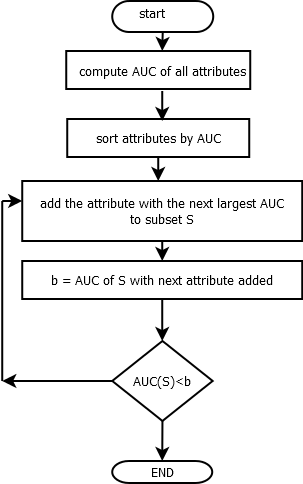}
	\caption{Rating scale stepwise reduction heuristic algorithm}
	\label{figure:rsr}
\end{figure}

In the \textbf{RatingScaleReduction}, the implemented algorithm (when reduced to its minimum) uses a loop for all attributes (with the class excluded) to compute AUC. Subsequently, attributes are sorted in the ascending order by AUC. The attribute with the largest AUC is added to a subset of all attributes (evidently, it cannot be empty since it is supposed to be the minimum subset S of all attributes with the maximum AUC). We continue adding the next in line (according to AUC) attribute to the subset S checking AUC. If it decreases, we stop the procedure. There are a lot of checking (e.g., if the dataset is not empty or full of replications) involved. These steps are implemented in \textit{startAuc}, \textit{totalAuc} oraz \textit{rsr} functions of the package.

Before running the RSR procedure the data set should be analysed to detect replicated examples and so-called "gray" examples.  One example may be replicated $m$ times, where $m$ is the total number of examples, so that there are no other examples. Such situation would deviate computations and should be early detected. Ideally, no example should be replicated but if the replication rate is small, we can proceed to computing AUC. There is no generally acceptable ``golden rule'' for the level or replication rate. Moreover the data may contain gray examples which should also be detected, gray example is an example for which there are another examples in the data set having identical values on all attributes but different decision. This analysis of data set can by carry out using functions: \textit{diffExamples} and \textit{grayExamplesN}, \textit{grayExamples}.

The important problem after the scale reduction by RSR procedure is to check for the possible inclusion of the next attribute in the reduced rating scale by maximizing AUC of all included items. In a highly unlikely scenario, all attributes will be included in the reduced (that is, non reduced) set of items. The reduced rating scale of one attribute may be created if there is an identifying attribute. To test the inclusion the function \textit{CheckAttr4Inclusion} is available in the package.

\section{RatingScaleReduction: overview of the package functions}
The \textbf{RatingScaleReduction} package implements the above-stated stepwise procedure using two functions of the \textbf{pROC} package: \textit{roc} and \textit{roc.test}. It works on the data as the \textit{matrix} or \textit{data.frame} containing columns of attributes and one decision column with two categories, e.g. (0,1). The rows in \textit{data.frame} represents examples in the sample. All attributes and the decision vector must be numeric. There are to groups of functions available in the package. Because the essence of the procedure is to set the attributes in the correct order good practice is to enter their name using e.g. \textit{colnames} in \textbf{R}. The first group are dedicated to carry out the RSR procedure:
\begin{enumerate}
\item{\textit{startAuc(attribute, D)} -- compute the AUC values of every single attribute in the rating scale.}

\item{\textit{totalAuc(attribute, D, plotT=FALSE)} -- sort AUC values in the ascending order and compute AUCs of running total of first $k$ attributes, $k=1,...,n$, where $n$ is the number of attributes. Setting the argument \textit{plotT} as \textit{TRUE} the plot of new AUC values is created. The horizontal line marks the max new AUC.}

\item{\textit{rsr(attribute, D, plotRSR=FALSE)} -- the main function of the package reducing the rating scale according the procedure illustrated by Fig.~\ref{figure:rsr}. Setting the argument \textit{plotRSR} as \textit{TRUE} the plot of ROC curve of the sum of attributes in reduced rating scale is created. }
\end{enumerate}
Additionally, the package provides second group of functions to support the reduction procedure:
\begin{enumerate}
\item{\textit{CheckAttr4Inclusion(attribute, D)} -- carry out a statistical tests for a difference in AUC of two correlated ROC curves: ROC1 of the sum of attributes from reduced rating scale and ROC2 of this sum plus the next ordered attribute. The function uses The function \textit{roc.test} from the \textbf{pROC} is used and all implemented tests are available, in particular \textit{delong} and \textit{bootstrap}.}

\item{\textit{diffExamples(attribute)} -- search replicated examples in the data and return the number of different examples and the number of duplicates. }

\item{\textit{grayExamples(attribute, D)} -- produce the list of pairs of examples having identical values on all attributes. The decision value and attributes are produced for every pair in the data set, so the list clearly shows all gray examples.}

\item{\textit{grayExamplesN(attribute, D, N)} -- produce a list of examples and the numbers of examples $j$ in the data set having identical values on all attributes for the given example $N$.}

\end{enumerate}

\subsection{Data sets used for verification and working with the package}

The examples presents the capabilities of the \textbf{RatingScaleReduction} package. The full code is available for download from \\ \textit{https://github.com/woali/RatingScaleReduction/example\_Rj.r}.

\subsection{The first demonstration example: rating scale reduction}
We consider the data BDI data set used in \cite{K17rsr}. It is a rating scale for depression BDI (Beck Depression Inventory) with 21 attributes in our relational database. The goal in this example is to show how to reduce the BDI rating scale in the use of three main functions of the package.
The \textit{data.frame} we work on contains 21 columns with attributes and one additional column as a decision (reality). The sample is represented  with 561 examples (instances).

We start with analysis of the AUC of all 21 individual attributes of BDI scale by the use of the function \textit{totalAuc} setting the argument \textit{plotT} as \textit{TRUE}.

\begin{verbatim}
> tauc.bdi <-totalAuc(attribute, D, plotT=TRUE)

> tauc.bdi$summary

            AUC one variable AUC running total
BDI_1         0.7250092         0.7250092
BDI_14        0.7074013         0.7765412
BDI_7         0.7014889         0.7945490
BDI_9         0.7004614         0.8095300
BDI_10        0.6972253         0.8131352
BDI_15        0.6920635         0.8221669
BDI_17        0.6742833         0.8205426
BDI_8         0.6679833         0.8192322
BDI_20        0.6669989         0.8198782
BDI_5         0.6584779         0.8211333
BDI_3         0.6556663         0.8210840
BDI_4         0.6511874         0.8211394
BDI_13        0.6489172         0.8210779
BDI_12        0.6367909         0.8195583
BDI_2         0.6312046         0.8186908
BDI_19        0.6292851         0.8181125
BDI_18        0.6100283         0.8160699
BDI_6         0.6100037         0.8138489
BDI_16        0.6059370         0.8116525
BDI_11        0.5973422         0.8110680
BDI_21        0.5874677         0.8117263

\end{verbatim}

The \textbf{R} output shows the \textit{tauc.bdi\$summary} AUC of every single attribute in the second column, sorted in the ascending order. The running total of AUCs is in the thrird column. The initially selected variable (BDI\_1) for the first row is the attribute with the largest AUC. Subsequently, we add to it the variable with the largest AUC of the remaining attributes. The process continues while the last attribute of the scale is added.

\begin{figure}[hbp!]
	\centering
	\includegraphics[scale=.2]{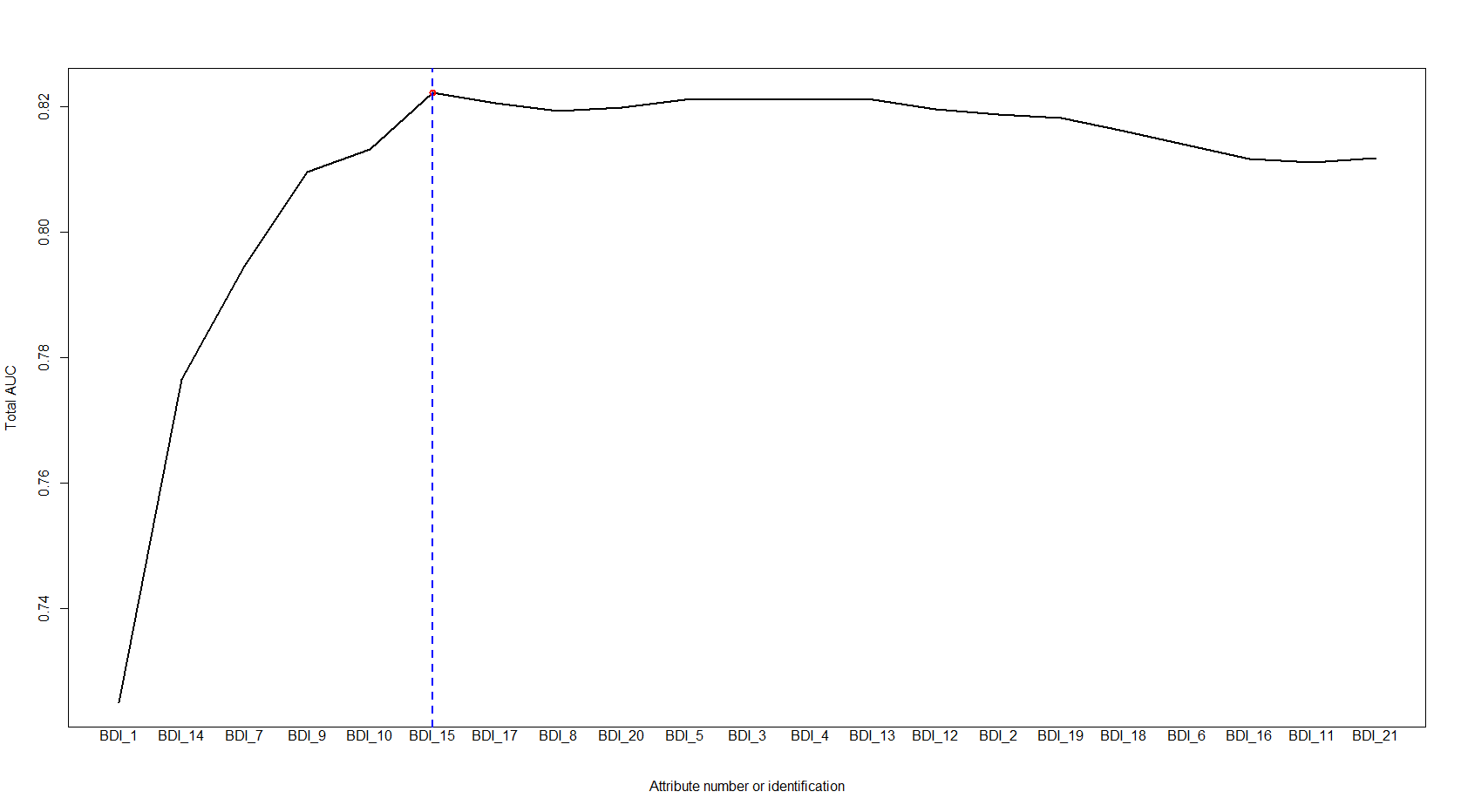}
	\caption{A stepwise AUC reduction method (example)}
	\label{figure:fig1}
\end{figure}

Values in the running total (from the top to the current variable) are checked for growing. Evidently, the value 0.725 in the first row is the same for the running total as for the single variable (BDI\_1). However, the value in the third row (0.795) is not for variable 7 but the total of variables BDI\_1, BDI\_14, and BDI\_7. In particular, the value (0.812) in in the last row is for the total of all variables. Their line plot can be easily created by setting the \textit{totalAuc} parameter \textit{plotT} as \textit{TRUE}. The plot for our BDI scale is illustrated by Fig.~\ref{figure:fig1}. The curve peek is for variable \#6 which is 15. \\

\newpage
Printing the value \textit{tauc.bdi\$item} we receive the attribute labels in an ascending order.

\begin{verbatim}

> tauc.bdi$item

"BDI_1"  "BDI_14" "BDI_7"  "BDI_9"  "BDI_10" "BDI_15"
"BDI_17" "BDI_8"  "BDI_20"
"BDI_5"  "BDI_3"  "BDI_4"  "BDI_13" "BDI_12" "BDI_2"
"BDI_19" "BDI_18" "BDI_6"
"BDI_16" "BDI_11" "BDI_21"
\end{verbatim}

As illustrated by Fig.~\ref{figure:fig1}, the value of AUC of the selected subset of attributes is increasing by adding the first six attributes labeled DBI\_1, DBI\_17, DBI\_7, DBI\_9, DBI\_10, and DBI\_15. There is a slight decline by adding the variable DBI\_16.
For this reason, the reduction procedure is terminated after the first six attributes are added.
The function \textit{rsr} reduce the scale automatically assuming the truncation point as the attribute that first reaches the maximum AUC. AUC is a real value between 0 and 1. It is 0.5 for random data but hardly ever reaches 1 since , in reality, there are always ``gray examples'' in sizable data.

\newpage
\begin{verbatim}

> rsr.bdi <-rsr(attribute, D, plotRSR=TRUE)
The criteria: Stop first MAX AUC
> rsr.bdi$rsr.auc

[1] 0.7250092 0.7765412 0.7945490 0.8095300 0.8131352 0.8221669

$rsr.label
[1] "BDI_1"  "BDI_14" "BDI_7"  "BDI_9"  "BDI_10" "BDI_15"

$summary
AUC one variable AUC running total
BDI_1         0.7250092         0.7250092
BDI_14        0.7074013         0.7765412
BDI_7         0.7014889         0.7945490
BDI_9         0.7004614         0.8095300
BDI_10        0.6972253         0.8131352
BDI_15        0.6920635         0.8221669

\end{verbatim}

\begin{figure}[htbp]
	\centering
	\includegraphics[scale=.5]{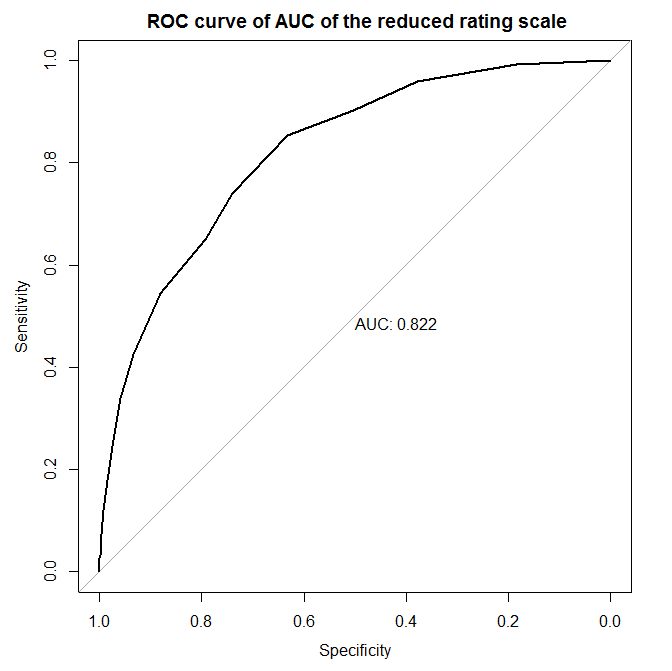}
	\caption{AUC of $BDI_1 + BDI_{14} + BDI_7 + BDI_9 + BDI_{10} + BDI_{15},$ for DBI scale }
	\label{figure:fig2}
\end{figure}

Setting the \textit{rsr} parameter \textit{plotRSR} as \textit{TRUE} the function generate plot illustrated by Fig.~\ref{figure:fig2}. \\

We assume that by selecting the ``best'' attribute in a loop, we are able to reduce the number of attributes for the best preventiveness. In our case, having the largest AUC is the ``best'' criterion. Adding the next ``best'' attribute to the selected attribute from the subset of the remaining attributes until AUC of all selected attributes decreases is the main idea of our heuristic. So far, each and every rating scale has been reduced.

\subsection{Second illustrative example: using the entire RSR procedure}
Le us consider the data set Hepatitis analyzed in \cite{MLrepository} and located at\\
\textit{http://archive.ics.uci.edu/ml/}.  
It has 20 attributes and 312 examples used in \cite{cestnik1987assistant}. The goal is to illustrate how our entire RSR procedure may be used. To reduce Hepatitis data set, we use the following attributes \textit{hepato}: \\

\begin{verbatim}

> names(att[,-d3])
 "time"     "status"   "trt"      "age"      "sex"
 "ascites"  "spiders"
 "edema"    "bili"     "chol"     "albumin"  "copper"
 "alk.phos" "ast"
  "trig"     "platelet" "protime"  "stage"
\end{verbatim}

\noindent The following steps are needed: \\

\begin{itemize}
\item{detect duplicates and gray example in data},
\item{reduce rating scale},
\item{check possible inclusion}.
\end{itemize}

\noindent By executing this code:\\

\begin{verbatim}

> diffExamples(att[,-d3])
$total.examples
[1] 276

$dif.examples
[1] 276

$dup.examples
[1] 0

\end{verbatim}

\noindent we detect no duplicate in data set.\\

Subsequently, we analyze the data set to detect gray examples for \textit{(status, sex, spiders, stage)} attributes.
Gray examples are located by the use of \textit{data.frame} and a function called \textit{gray.ex}.

Working on a full data set is a time-consuming process since it requires all pair comparisons to be analyzed.
A short optimization procedure is used for attributes in two categories.
The code below shows how to list by gray examples by comparing the subset of attributes \textit{(status, sex, spiders, stage)}
using the function \textit{grayEcamplesN}. The key issue is to properly modify the data.

\newpage
\begin{verbatim}

> gray.ex <-c()
> df1 <-unique(data.frame(hepato, status, sex, spiders, stage))

> for (i in 1:nrow(df1)){
> ex <-grayExamplesN(df1[,2:ncol(df1)], df1[,1], i)$examp
> if (nrow(ex)>1){
> gray.ex <-rbind(gray.ex, ex[1,])}}
> colnames(gray.ex) <-names(df1)

> gray.ex
    hepato status sex spiders stage
1        1      0   1       1     4
2        1      0   1       1     3
3        0      0   2       0     4
6        1      0   1       0     3
7        0      0   1       0     3
8        0      0   1       1     2
9        0      0   1       1     4
15       1      0   1       0     4
22       1      0   2       0     2
23       0      0   1       0     2
29       1      0   1       0     2
42       0      0   2       0     3
48       1      0   1       1     2
53       0      0   1       0     4
54       0      0   1       1     3
57       1      0   2       0     3
70       1      0   2       0     4
80       0      0   2       0     2
93       1      1   1       0     3
98       0      1   1       0     3
111      1      1   1       0     4
139      0      1   1       0     2
253      1      1   1       0     2
256      0      1   1       0     4

\end{verbatim}
\noindent The \textbf{R} print the list of gray examples. It means that every example from list corresponds to (not listed) examples having identical attributes, but different decision.

To reduce this rating scale, we execute the functions: \textit{totalAuc} and \textit{rsr}.\\

\begin{verbatim}

> totalAuc(att[,-d3], hepato, plotT=TRUE)
> rsr(att[,-d3], hepato, plotRSR=TRUE)

\end{verbatim}

\begin{figure}[htbp]
	\centering
	\includegraphics[scale=.2]{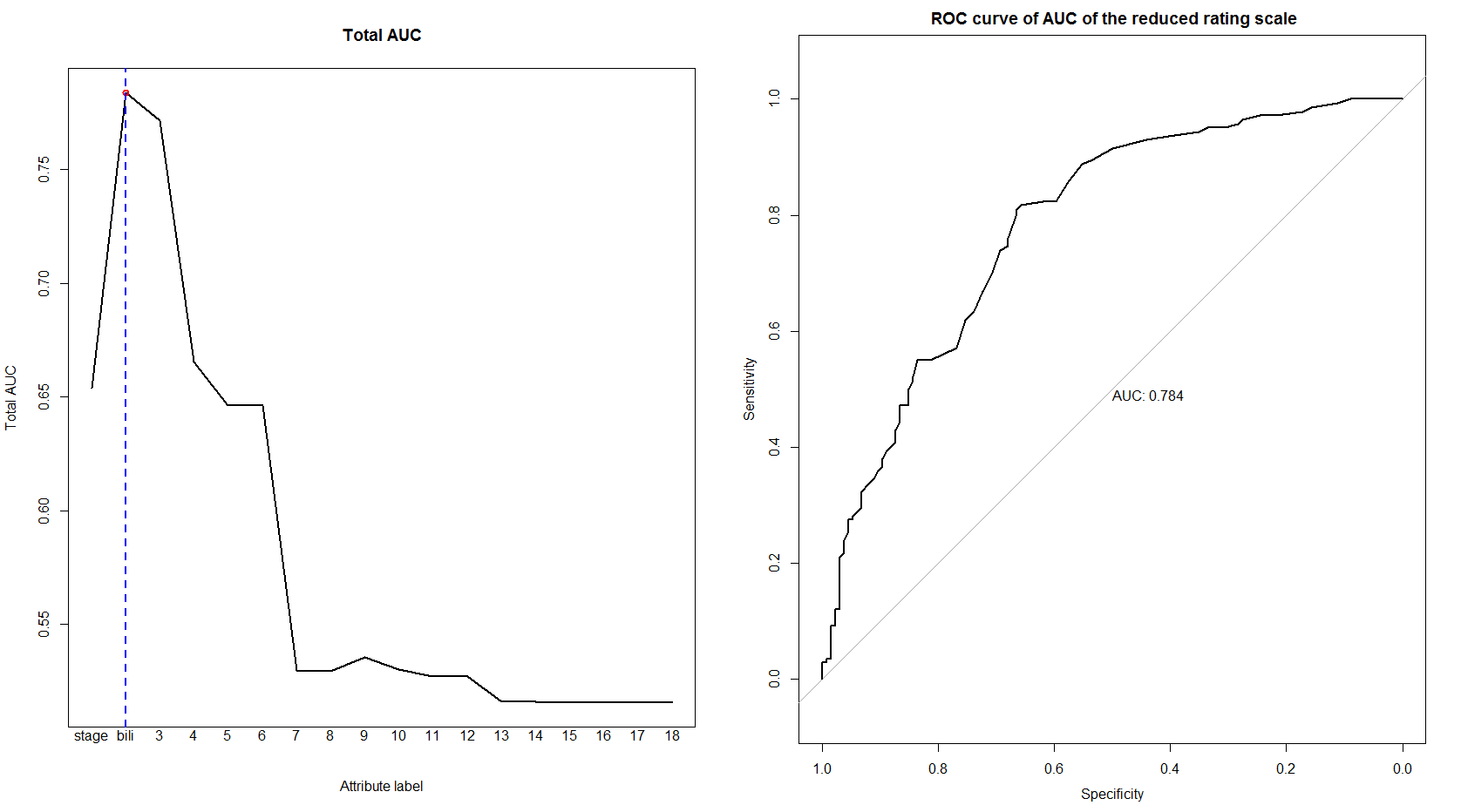}
	\caption{Left panel: AUC of individual attributes of \textit{Hepatitis} data set; Right panel: AUC of \textit{stage+bili} for \textit{hepato} scale}
	\label{figure:fig3}
\end{figure}

After reduction, the scale contains two attributes: \textit{stage, bili}. The plot in Fig.~\ref{figure:fig3} illustrates AUC of the reduced rating scale and the proper ROC curve.\\

Finally, we examine the scale for possible inclusion using deLong and boostrap tests as in the function \textit{roc.test} from the \textbf{pROC}.

\begin{figure}[h]
	\centering
	\includegraphics[scale=.5]{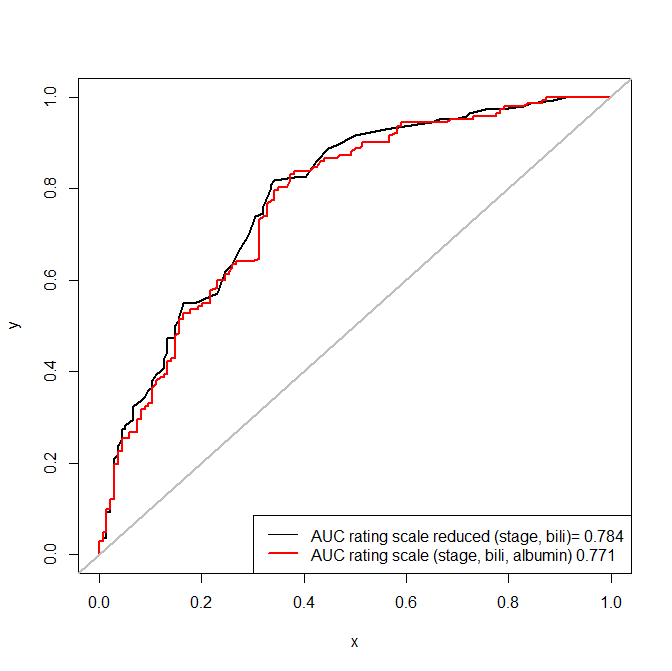}
	\caption{ROC curves for the original and reduced rating scales for the Hepatitis data set}
	\label{figure:fig4}
\end{figure}

\newpage
\begin{verbatim}

> ROC1 <-CheckAttr4Inclusion(att[,-d3], hepato,
+ method=c("delong"), alternative=c("two.side"))

> ROC2 <-CheckAttr4Inclusion(att[,-d3], hepato,
+ method=c("bootstrap"), alternative=c("two.side"))

> summ
          Z statistics    p-value
DeLong        1.986207 0.04701039
bootstrap     1.945672 0.05169413

\end{verbatim}
\noindent The \textit{p-value = 0.04701} shows that according to \textit{deLong} test the null hypothesis ``\textit{H0}: true difference in AUC is equal to 0'' should not be rejected in favor of the alternative. Fig.~\ref{figure:fig4} illustrates two tested ROC curves.

\section{The potential application targets}

Rating scales are by far more important contributors to science that we can address them by this study. Most examinations for granting scientific degrees are rating scales of various  shapes and forms. Simplifying them (or reducing in size) is needed since we are subjected to more and more examination for so needed certifications.

In bioinformatics, reporting trade-off in sensitivity and specificity, by using a Receiver Operating Characteristic (ROC) curve, is becoming a common practice.
ROC plot has the sensitivity on the $y$ axis, against the false discovery rate (1- specificity) on the $x$ axis.
ROC curve plot provides a visual tool to determine the boundary limit (or the separation threshold)
of a subset (or a combination) of scale items for the potentially optimal combination
of sensitivity and specificity. The area under the curve (AUC) of the ROC curve indicates
the overall accuracy and the separation performance of the rating scale.
It can be readily used to compare different item subsets.
As a rule of thumb, the fewer the scale items used to maximize the AUC of the ROC curve, the better.

World Health Organization estimates are included behind selected rating scales for mental disorder.
Rating scales are of considerable importance for psychiatry where they are predominately used for screening patients for mental disorders such as: \\
\begin{itemize}
\item depression (see \cite{K17rsr}) which affects 60 million people worldwide according to \cite{WHOfs},
\item bipolar affective disorder (60 million people),
\item dementia and cognitive impairment (47.5 million people)
\item schizophrenia (21 million people),
\item autism and autism spectrum disorder(e.g., \cite{KKW2012})
\item mania and bipolar disorder,
\item addiction,
\item personality and personality disorders,
\item anxiety,
\item ADHD;
\end{itemize}
and many other disorders.

Usually, there are many scales for each mental disorder. The most important for screening are global scales. Reducing these global rating scales makes they more usable as indicated in \cite{K17rsr}.
World Health Organization Media Centre reports that``depression and anxiety disorders cost the global economy US \$1 trillion each year'' and it is no longer a local problem.

\section{Conclusions}

The presented method has reduced the number of the rating scale items (variables) to 28.57\% from the original number of items (from 21 to 6). It means that over 70\% of collected data was unnecessary. It is not only an essential budgetary saving, as the data collection is usually expensive and may easily go into hundreds of thousands of dollars, but the excessive data collection may contribute to the data collection error increase. The more data are collected, the more errors may occur occur since a lack of concentration and boredom are realistic factors.

By using the proposed AUC ROC reduction method, the predictability has increased by approximately 0.5\%. It may seem insignificant. However, for a large population, it is of considerable importance. In fact, \cite{WHOfs} states that: ``Taken together, mental, neurological and substance use disorders exact a high toll, accounting for 13\% of the total global burden.''

The proposed use of AUC for reducing the number of rating scale items, as a criterion, is innovative and applicable to practically all rating scales. System R code is posted on the Internet (RatingScaleReduction) for the general use as a R package. Certainly, more validation cases would be helpful and the assistance will be provided to anyone who wishes to try this method using his/her data.

Future plans include using the presented method for financial data analyzed 
but the real aim our collaborative effort is towards psychiatric scales. The reduced scales can be further enhanced by the method described in \cite{KKW2012} and \cite{K1996}.

\section*{Acknowledgments}
The first author has been supported in part by the Euro Research grant ``Human Capital''.

The authors would also like to express appreciation to Amanda Dion-Groleau, B.A. Honors Psycholgie, (Laurentian University, Psychology),
Tyler D. Jessup (Laurentian University, Computer Science), and Grant O. Duncan, Team Lead, Business Intelligence, Integration and Development, Health Sciences North, Sudbury, Ontario, Canada) for the editorial improvements.

\bibliographystyle{plain}
\bibliography{koczkodaj-wolny}

\begin{thebibliography}{10}

\bibitem{cestnik1987assistant}
Bojan Cestnik, Igor Kononenko, Ivan Bratko, et~al.
\newblock Assistant 86: A knowledge-elicitation tool for sophisticated users.
\newblock In {\em EWSL}, pages 31--45, 1987.

\bibitem{TF2004}
Tom Fawcett.
\newblock Roc graphs: Notes and practical considerations for researchers.
\newblock {\em Machine learning}, 31(1):1--38, 2004.

\bibitem{goksuluk2016easyroc}
Dincer Goksuluk, Selcuk Korkmaz, Gokmen Zararsiz, and A~Ergun Karaagaoglu.
\newblock easyroc: An interactive web-tool for roc curve analysis using r
  language environment.
\newblock {\em The R Journal}, 2016.

\bibitem{OHQ}
Peter Hills and Michael Argyle.
\newblock The oxford happiness questionnaire: A compact scale for the
  measurement of psychological well-being.
\newblock {\em Personality and individual differences}, 33(7):1073--1082, 2002.

\bibitem{KKW2012}
Tamar Kakiashvili, Waldemar~W. Koczkodaj, and Marc Woodbury{-}Smith.
\newblock Improving the medical scale predictability by the pairwise
  comparisons method: Evidence from a clinical data study.
\newblock {\em Computer Methods and Programs in Biomedicine}, 105(3):210--216,
  2012.

\bibitem{K1996}
Waldemar~W Koczkodaj.
\newblock Statistically accurate evidence of improved error rate by pairwise
  comparisons.
\newblock {\em Perceptual and Motor Skills}, 82(1):43--48, 1996.

\bibitem{K17rsr}
W.W. Koczkodaj, T.~Kakiashvili, A.~Szymańska, J.~Montero-Marin, R.~Araya, ,
  J.~Garcia-Campayo, K.~Rutkowski, and D.~Strzałka.
\newblock How to reduce the number of rating scale items without predictability
  loss?
\newblock {\em Scientometrics}, 2, 2017.

\bibitem{leeper2016crowdsourced}
Thomas~J Leeper.
\newblock Crowdsourced data preprocessing with r and amazon mechanical turk.
\newblock {\em The R Journal}, 8(1):276--288, 2016.

\bibitem{MLrepository}
M.~Lichman.
\newblock {UCI} machine learning repository, 2013.

\bibitem{MBB2002}
Roger McIntyre, Sidney Kennedy, Michael Bagby, and David Bakish.
\newblock Assessing full remission.
\newblock {\em Journal of psychiatry \& neuroscience: JPN}, 27(4):235, 2002.

\bibitem{pROC}
Xavier Robin, Natacha Turck, Alexandre Hainard, Natalia Tiberti, Frédérique
  Lisacek, Jean-Charles Sanchez, and Markus Müller.
\newblock proc: an open-source package for r and s+ to analyze and compare roc
  curves.
\newblock {\em BMC Bioinformatics}, 12:77, 2011.

\bibitem{sing2005rocr}
Tobias Sing, Oliver Sander, Niko Beerenwinkel, and Thomas Lengauer.
\newblock Rocr: visualizing classifier performance in r.
\newblock {\em Bioinformatics}, 21(20):3940--3941, 2005.

\bibitem{VC1979}
W.~Velden, M. \&~Clark.
\newblock Reduction of rating scale data by means of signal detection theory.
\newblock {\em Psychophysics}, 25(6):517--518, 1979.
\newblock doi:10.3758/BF03213831.

\bibitem{WHOfs}
WHO.
\newblock Mental disorders fact sheet.
\newblock {\em ``www.who.int/mediacentre/factsheets/fs396/en/''}, April, 2016.

\bibitem{winslow2016sbtools}
Luke~A Winslow, Scott Chamberlain, Alison~P Appling, and Jordan~S Read.
\newblock sbtools: A package connecting r to cloud-based data for collaborative
  online research.
\newblock {\em The R Journal}, 8(1):387--398, 2016.

\end{thebibliography}

\bigskip

\noindent Waldemar W. Koczkodaj\\
Computer Science\\
Ramsey Lk Rd.\\
Laurentian University \\
Sudbury, Ontario P3E 2C6 \\
Canada \\
\textmd{wkoczkodaj@cs.laurentian.ca}\\
\textmd{http://cs.laurentian.ca/wkoczkodaj/info}\\

\medskip
\noindent Alicja Wolny--Dominiak\\
	Department of Statistical and Mathematical Methods in Economics \\
	University of Economics in Katowice\\
	ul. 1 Maja 50\\
	40--287 Katowice\\
	Poland\\
\textmd{alicja.wolny-dominiak@uekat.pl} \\
\textmd{web.ue.katowice.pl/woali}\\

\end{document}